\documentclass[prl,amsmath,amssymb,superscriptaddress,twocolumn]{revtex4}
\usepackage{graphicx}

\newcommand{\binomial}[2]{\genfrac{(}{)}{0pt}{}{#1}{#2}}
\renewcommand{\Psi}{\varPsi}

\begin{document}

\title{Discreteness and the origin of probability in quantum mechanics}

\author{Roman~V.~Buniy} \email{roman@uoregon.edu}
\affiliation{Institute of Theoretical Science, University of Oregon,
Eugene, OR 97403}

\author{Stephen~D.~H.~Hsu} \email{hsu@duende.uoregon.edu}
\affiliation{Institute of Theoretical Science, University of Oregon,
Eugene, OR 97403}

\author{A.~Zee}
\email{zee@itp.ucsb.edu}
\affiliation{Kavli Institute for Theoretical Physics,
UCSB, Santa Barbara, CA 93106}

\begin{abstract}
Attempts to derive the Born rule, either in the Many Worlds or
Copenhagen interpretation, are unsatisfactory for systems with only a
finite number of degrees of freedom. In the case of Many Worlds this
is a serious problem, since its goal is to account for apparent
collapse phenomena, including the Born rule for probabilities,
assuming only unitary evolution of the wavefunction. For finite number
of degrees of freedom, observers on the vast majority of branches
would not deduce the Born rule. However, discreteness of the quantum
state space, even if extremely tiny, may restore the validity of the
usual arguments.
\end{abstract}



\maketitle

\section{Introduction: the problem with probability} 

Quantum mechanics exhibits an odd dichotomy in the time evolution of
states. A quantum state undergoes deterministic, unitary evolution
until a measurement causes probabilistic, non-unitary collapse. While
many physicists do not feel that there is anything wrong with this
standard Copenhagen picture, it seems less than economical to
postulate two fundamental processes---unitary evolution and
non-unitary measurement---if somehow one could suffice. Everett
\cite{Everett} proposed that unitary time evolution of a closed system
is sufficient to account for the \emph{appearance} of measurement
collapse to observers inside the system (see also Hartle \cite{Hartle}
and DeWitt and Graham \cite{DeWitt}), in what has now become known as
the Many Worlds (MW) formulation of quantum mechanics.

The MW interpretation is regarded as extravagant, and hence
implausible, by many (including at least one of the authors), because
of the huge multiplicity of branches of the wavefunction, each of
which is presumed to be as real as the others \cite{fn1}.  Before the
anti-MW reader abandons this paper, we note that the discussion that
follows applies also to the conventional Copenhagen interpretation,
with measurement collapse, and may allow a derivation of probability
in quantum mechanics from a weaker initial assumption, known as the
certainty assumption, along the lines of Hartle \cite{Hartle} (see
also Farhi, Goldstone and Gutmann \cite{Farhi} and Coleman and
Lesniewski \cite{CL}). An attractive doctrine (preferred by one of the
authors) is the minimalist view outlined by Hartle \cite{Hartle2}
insisting that physics should be done without ill-defined words and
slogans such as ``The other worlds are just as real." Our analysis could
also be read within this post-Everett or decoherent histories
approach.

We focus on the Born rule in quantum
mechanics, and the extent to which it can be derived. The Born rule
states that given an observable $A$ with spectrum $\lambda_i$
and eigenstates $\vert \psi_i \rangle$, the probability of $\lambda_i$
as the outcome of a measurement on state $\vert \psi \rangle$ is $P_i
= \vert \langle \psi_i \vert \psi \rangle \vert^2$. \emph{It has been
claimed by Everett, Hartle, and others, that this rule arises as a
consequence of the assumption of unitary evolution, but as we discuss
below, the derivation is unsatisfactory for any system
with only a finite number of degrees of freedom.} (For recent
discussions of the Born rule in MW, see \cite{others}.)

In a recent paper \cite{BHZ} we speculated that quantum gravity and
related considerations may imply that quantum state space is itself
discrete. We will review our argument in the next section. Here
we point out that one consequence of this discreteness in state space
may be the emergence of the Born rule, even in the case when the
number of degrees of freedom is finite.

The original derivation of the Born rule given by Everett
\cite{Everett}, Hartle \cite{Hartle}, and others, is quite simple.
Consider an ensemble of identically prepared states
\begin{equation}
\Psi  = \psi  \otimes \cdots \otimes  \psi  = \otimes_{a=1}^N ~ \psi^{(a)} ,
\end{equation}
and a sequence of outcomes $S = ( s_1, s_2, \ldots , s_N )$ obtained
from measurements on each of the states. The probability $P(S)$ of a
given sequence, or class of sequences, calculated using the Born rule,
is identical to the norm (magnitude) squared of the projection of
$\Psi$ onto eigenstates with the eigenvalues $( s_1, s_2, \ldots,
s_N)$, namely $\vert \langle s_1 s_2 \ldots s_N \vert \Psi \rangle
\vert^2$. As Everett noted, it follows that an improbable sequence
corresponds to a component of $\Psi$ (in the eigenstate basis) with
small magnitude. In the formal limit $N \rightarrow \infty$,
components of $\Psi$ which do not correspond to statistically typical
sequences generated by the Born rule have zero magnitude
(i.e. converge to the null vector), and therefore do not correspond to
physical states. From the frequentist perspective on probability,
then, the Born rule is a consequence of excluding zero norm states
from the Hilbert space.

To further elucidate, consider a simple example using spin states. Let
$\vert \psi \rangle = c_+ \vert + \rangle + c_- \vert - \rangle$, and
define $p_\pm = \vert c_\pm \vert^2$. Then a sequence of measurement
outcomes will be of the form $S = \{ + + - + \cdots \}$. If the
sequence is generated by the Born rule, then in the limit of large
$N$, the fraction of $(+)$ outcomes will be $p_+$ to very good
approximation. Any other value for the fraction of $(+)$ outcomes has
zero probability at infinite $N$. Correspondingly, the magnitude squared
$\vert \langle s_1 s_2 \cdots s_N \vert \Psi \rangle \vert^2$ is
zero for any state $\langle s_1 s_2 \cdots s_N \vert$ in
which the fraction of outcomes $s_i$ equal to $(+)$ is not $p_+$.

This can be generalized: if $\langle s_1 s_2 \cdots s_N \vert$
corresponds to a sequence $S = \{ s_1, s_2, \ldots, s_N\}$ which is
statistically atypical according to the Born rule, its overlap with
$\Psi$ will vanish when $N \rightarrow \infty$. Everett referred to
these branches of the wavefunction as ``maverick worlds''---observers
on these branches would not deduce the Born rule. Below, we will
repeat this discussion for those readers who prefer a more standard
Copenhagen interpretation to the MW interpretation.

We can define
parameters characterizing the deviation of a maverick world from the
central Born value. For example, in the spin example, we might
consider $f_+ $ to be the frequency of $(+)$ outcomes, so that $\delta
= f_+ - p_+$ is the deviation parameter. Then any branch with non-zero
$\delta$ will have vanishing norm in the large $N$ limit. When $N$ is
strictly infinite all maverick worlds have zero norm. The remaining
branches have outcomes $S$ which satisfy the Born rule in the
frequentist sense.

The problem with this reasoning is of course that $N$ is never
strictly infinite. In fact, given the finite size of the causal
horizon of our universe and an ultraviolet cutoff on modes (e.g.,
from the Planck scale), we obtain a finite, although very large, upper
limit on the number of outcomes $N$ which characterize any particular
branch of the MW wavefunction. Without invoking something like the
Born rule---a correspondence between probability and norm---there
is no reason to exclude branches with small but non-zero norm. The
problem is exacerbated by the fact that maverick worlds are generally
far more numerous than non-maverick worlds. The MW wavefunction
branches with each measurement, regardless of how small either of
$\vert c_\pm \vert^2$ is. This leads to $2^N$ total branches after $N$
measurements. Even if, e.g., $\vert c_+ \vert^2$ is much larger than
$\vert c_- \vert^2$, both $(+)$ and $(-)$ outcomes will still occur at
each branch, and the structure of the tree is independent of $c_\pm$
as long as neither is zero. The overwhelming majority of branches will
have roughly equal numbers of $(+)$ and $(-)$ outcomes. Thus the
multiplicity of maverick worlds is enormously larger than non-maverick
worlds, although their collective magnitude is vanishingly small. Again,
without assuming the Born rule, we have no \textit{\`a priori} reason
to exclude small (but non-zero) norm states.

Of course, a strict frequentist interpretation of probability requires
an infinite sequence of outcomes. However, the use of probability by
physicists is more Bayesian than frequentist: confronted with a
\emph{finite} sequence of outcomes, $S = ( s_1, s_2, \ldots , s_N )$,
our goal is to deduce a predictive model for subsequent outcomes. In
this way, we deduce the Born rule based on the limited number of
measurements thus far performed on quantum systems.
 
As mentioned, our discussion may be of interest even to those who do
not accept MW, as it pertains to the origin of the Born rule within
the Copenhagen, or measurement collapse, interpretation. In
particular, it has been proposed by Hartle \cite{Hartle} that the Born
rule can be derived from the weaker \emph{certainty assumption},
stating that when a measurement of an observable $A$ is performed on
an eigenstate $\vert a \rangle$ of $A$, the value $a$ is obtained with
certainty. Taking $A$ to be, for example, the frequency operator for
$(+)$ outcomes, or any other statistical property, Hartle found that
for $N$ infinite, $\Psi$ is an eigenstate of each of these
statistical operators, with eigenvalues given by the Born rule.

The discussion parallels that in the MW interpretation. In the
standard Copenhagen picture the state $\Psi$ is, in the eigenstate
basis, a sum of $2^N$ terms, each term being in one-to-one
correspondence with a MW branch or a universe. In the Copenhagen
interpretation the outcomes $S$ result from measurements on an
ensemble, whereas in MW they specify a particular branch or decoherent
history \cite{dh} of the wavefunction of the entire universe. The
mathematics is the same in either picture: maverick terms collectively
have a very small norm that approaches zero as $N$ approaches
infinity.

This has the same weakness as the earlier MW argument. For any finite
$N$, the state $\Psi$ is only approximately an eigenstate of the
frequency operator. The certainty assumption does not specify the
outcome of a measurement on an approximate eigenstate, and going
further requires an assumption relating the norm of a state vector to
the probability of a measurement outcome, which is essentially the
Born rule.

\section{Discrete state space} 

Consider normalized states $\Psi = \psi \otimes \cdots \otimes
\psi$ and $\Psi' = \psi' \otimes \cdots \otimes \psi' $. Suppose that,
due to fundamental discreteness, one cannot distinguish $\psi$ and
$\psi'$ when $\vert \, \psi - \psi' \vert < \epsilon$. This implies
that the direct product states cannot be distinguished when (assuming
$\sqrt{N} \epsilon \ll 1$)
\begin{equation}
\label{dpsi}
\vert \, \Psi  -  \Psi'  \vert < \sqrt{N} \epsilon .
\end{equation} 
(We have assumed that $\langle \psi \vert \psi' \rangle$ is real,
which would be the case if $\psi'$ resulted from rotating $\psi$
slightly on the Bloch sphere. Relative phases could lead to order $N
\epsilon$ terms in Eq.~(\ref{dpsi}), which allow an acceptable cutoff
of maverick branches for even smaller discreteness scale $\epsilon$.)
Motivated by this observation, \emph{we assume that any (maverick!)
components of $\Psi$ with norm less than $\sqrt{N} \epsilon$ can be
removed from the wavefunction.}

\begin{figure}[h]
\includegraphics[width=4cm]{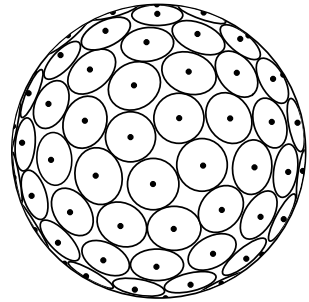}
\caption{A possible discretization of the Bloch sphere (qubit state
  space). Points on each disc (of size $\epsilon$) are
  identified. Points between discs can be assigned to the nearest disc.
}
\label{figure}
\end{figure}

We argued in Ref.~\cite{BHZ} that quantum gravity
suggests a discreteness scale of order $\epsilon \sim E$, where $E$ is
the characteristic energy of the system described by $\psi$, in Planck
units. Equivalently, $\epsilon \sim L^{-1}$, where $L$ is the
characteristic size, or Compton wavelength, of the system. We can
motivate this result by noting that quantum gravity seems to imply a
minimal length \cite{minlength} of order the Planck length. A minimal
length restricts our ability to distinguish two different orientations
of an experimental apparatus, such as a Stern-Gerlach device for
measuring the orientation of a spin. (Rotation of the device by an
angle less than $L^{-1}$ does not displace any component by more than
the Planck length.) Thus, the resulting ambiguity in the spin state
even after an ideal measurement is at least of order $\epsilon$ given
above (see Fig.~\ref{figure}). There is no way to ensure that the
ensemble states $\psi$ are identical to accuracy better than
$\epsilon$. For example, each time we pass a spin through the
Stern-Gerlach device to produce another $\psi$ there can be no
guarantee that the Stern-Gerlach device remains in precisely the same
orientation.

While some might consider fundamental discreteness of the space of
quantum states (previously referred to in the earlier paper \cite{BHZ} as discrete Hilbert space \cite{fn2}) to be a radical notion,
we find asserting its absolute continuity in the absence of any
supporting experimental evidence to be perhaps just as
speculative. Consider the case of spacetime: few would claim that
spacetime must be absolutely continuous (in fact, most likely it is
not \cite{minlength}); why should quantum state space be different?

It is worth emphasizing that the discreteness we propose has nothing
to do with the dimensionality of state space. Rather, it has to do with
whether the coefficients $c_i$ in an eigenstate 
expansion $\vert \psi \rangle = \sum_i c_i \vert i \rangle$ are
continuous or can only take on a discrete set of values (see Fig.~1).

We have not specified the concrete realization of discreteness, other
than to assume that states can be defined only modulo some fundamental
uncertainty. There are many ways to define the
evolution of a state in a discrete state space.  One method
would be to write the time evolution operator $e^{-iHt}$ as a product
of discrete evolution operators $e^{-i H \Delta t}$ and apply this
product of operators sequentially to the state, followed by the ``snap
to'' rule (``snap to nearest lattice site''; see Fig.~\ref{figure})
after each step.  This is equivalent to taking classical digital
computer simulations literally. That is, by accepting the finite
precision of the variable $\psi (x)$ in an ordinary computer program,
one obtains a naive discretization of Hilbert space with the ``snap
to'' rule implemented by simple numerical rounding. \emph{With limited
numerical precision, branches of the wavefunction with very small norm
are eventually discarded.}  This scheme leads to small violations of
linear superposition, but only at the level of $\epsilon$.

Interestingly, for $\epsilon \sim L^{-1}$, the condition that
discreteness have only a small effect on $\Psi$, $\sqrt{N} \epsilon
\ll 1$, leads to a condition on the number of degrees of freedom
reminiscent of holography \cite{h1}:
\begin{equation}
N \ll L^2 \sim A ,
\end{equation}
where $A$ is the surface area of the region. This bound implies far
fewer degrees of freedom than the usual extensive scaling $N \sim
L^3$. It can be deduced as a constraint from gravitational collapse
\cite{h2}. Excluding states from the Hilbert space of the $L^3$ volume 
which would have already caused gravitational
collapse to a black hole, we find the stronger condition
$N < A^{3/4} \sim L^{3/2}$.

\section{No maverick worlds} 

Consider the spin example from the first section. Let $n = n_+ = f_+
N$ be the number of $(+)$ outcomes in the sequence $S$. We suppress
the $+$ subscript in what follows. For $N\gg 1$, the function
\begin{equation}
P(n)=\binomial{N}{n} \, p^n(1-p)^{N-n}.
\end{equation}
has a sharp maximum at $n= p N$ and rapidly decreases for $n$
sufficiently far from it. The maximum results from a competition
between the combinatorial factor (multiplicity), which is peaked at
$n=N/2$, and the product $p^{n}(1-p)^{N-n}$, which is peaked at either
$n=0$ or $n=N$, unless $p_-$ is extremely close to $p_+$.  It follows
that when calculating $P(n)$ for $n$ not too far from $p N$, we make a
negligible error by assuming $n\gg 1$ and $N-n\gg 1$. The Stirling
formula gives
\begin{equation}
  P(fN)\approx \left[2\pi Nf(1-f)\right]^{-1/2} \exp\left[-N\phi(f)
  \right],
\end{equation}
where 
\begin{equation}
  \phi(f)=f\ln{(f/p)}+(1-f)\ln{[(1-f)/(1-p)]}
\end{equation}
and $f=n/N$. For large $N$ this becomes sharply peaked. Expanding
$\phi(f)$ around $f=p$, we find
\begin{equation}
  P(fN)\approx \left[2\pi Np(1-p)\right]^{-1/2}
  \exp\left[-\frac{N(f-p)^2}{2p(1-p)}\right].
\end{equation}
The collective magnitude squared of all maverick states 
$\vert \delta, N \rangle$
with frequency
deviation $\vert \delta \vert = \vert f - p \vert$ greater than
$\delta_0$ is
\begin{equation}
\sum_{\vert \delta \vert> \delta_0} \langle \delta, N \vert \delta, N
\rangle \approx 2N \int_{p + \delta_0}^\infty df~ P( f N)~.
\end{equation} 
One contribution to the sum comes from the range $f \in
[0,p-\delta_0]$ and the other from the range $f \in
[p+\delta_0,1]$. Note that we have replaced $f(1-f)$ in the overall
factor in $ P(fN)$ by $p(1-p)$. The resulting error should be
negligible for our purposes here.

Requiring that this collective magnitude squared is less than $N \epsilon^2$ yields 
\begin{equation}
\label{d0}
\delta_0 > N^{-1/2}\left[ 2p (1-p) \vert \ln (N \epsilon^2) \vert
\right]^{1/2}.
\end{equation} 
The maximum deviation $\delta$ for undiscarded branches vanishes as $N
\rightarrow \infty$ for fixed $p,\epsilon$. If, for finite $N$, an
experimenter could measure all $N$ outcomes which define his branch of
the wavefunction, he might find a deviation from the predicted Born
frequency $f = p$ as large as $\vert \ln ( N \epsilon^2 ) \vert^{1/2}$
standard deviations (i.e., measuring the deviation in units of
$N^{-1/2}$). Note that we are working in the regime $N\epsilon^2 \ll
1$. If the discussion in Ref.~\cite{BHZ} offers a valid guide, the number
$\epsilon$ may be much smaller than $10^{-20}$, so that even if $N$ is as
large as Avogadro's number, $N\epsilon^2$ will still be a small
number (see example below).
 
However, an experimenter is unlikely to be able to measure more than a
small fraction of the outcomes that determine his branch. Recall that
in MW a particular branch of the wavefunction is specified by the
sequence of outcomes $S = (s_1, s_2, \ldots, s_N)$. $N$ is the
\emph{total number} of decoherent outcomes on a branch, so it is
typically enormous---at least Avogadro's number if the system
contains macroscopic objects such as an experimenter. The experimental
outcomes available to test Born's rule will be a much smaller
number $N_* \ll N$ corresponding to a subset of the $s_i$ directly
related to the experiment. Any deviation from the Born rule of order
$N^{-1/2}$ will be well within the experimental statistical error of
order $N_*^{-1/2}$. Therefore the Born rule will be observed to hold
in all the branches which remain after truncation due to discreteness.
This would, however, not be true if we were to set $\epsilon$ to zero,
in which case $\vert \ln ( N \epsilon^2 ) \vert^{1/2}$ would be
infinite.

For definiteness, consider the following numerical example. Let the
discreteness scale be truly tiny: $\epsilon \sim 10^{-100}$, and let
$N \sim 10^{160}$, which is the Hubble four-volume in fermis. Then
$\vert \ln ( N \epsilon^2 ) \vert^{1/2} \sim 10$, so unless
experimenters can measure more than $10^{-2} N \sim 10^{158}$ quantum
outcomes, they will have insufficient statistics to exclude any of the
maverick branches which remain after truncation.

\section{Copenhagen again}

If we assume the Copenhagen (collapse) interpretation, our analysis
describes when the Born rule can be supplanted by the weaker
assumption of certainty of measurement outcome when the measured state
is an eigenstate. In a discrete Hilbert space it is natural to extend
the notion of eigenstate, so that states within the discreteness
distance $\epsilon$ of an eigenstate will also be considered
eigenstates. (More precisely, we cannot distinguish between any two
such states.) As discussed in the previous section, for large (but
finite) $N$, $\Psi$ is approximately an eigenstate of any statistical
operator (such as the frequency operator, but also higher moments)
with eigenvalue equal to the Born rule value. For example, the
wavefunction is sharply peaked at the Born rule frequency value of $f
= p$. If, motivated by the discreteness scale $\epsilon$, we simply
modify the certainty assumption to include states which are
approximate eigenstates, we will have deduced the Born rule from a
more elementary assumption.

There is, however, a
technical difficulty in defining how close a state $\Psi$ is to
being an eigenstate of an operator such as the frequency
operator. It would be natural to impose a certainty criteria as
follows. Given  $\Psi$ satisfying 
\begin{equation}
\label{cd}
\vert \Psi - \Psi_{f} \vert < \sqrt{N} \epsilon ~,
\end{equation}
where $\Psi_{f}$ is an eigenstate of the frequency operator with
eigenvalue $f$, we identify $\Psi$ with $\Psi_{f}$ and require that a
measurement of the frequency on $\Psi$ return the value $f$ with
certainty. The problem arises because, for finite $N$, no choice of
$\Psi = \otimes_{a=1}^N \psi^{(a)}$ is an exact eigenstate of the frequency
operator (except in the trivial cases where $\psi$ is already an
eigenstate such as $\vert + \rangle$ or $\vert - \rangle$, and in
those cases $f$ is either zero or one).  The state $\Psi_{f}$ does not
exist, except in the limit $N \rightarrow \infty$, so the distance
criteria in Eq.~(\ref{cd}) cannot be defined.  ($\Psi$ and $\Psi_f$ live
in Hilbert spaces of very different dimensions.)  One has to rely on
some other criterion for identifying a state $\Psi$
as a frequency eigenstate. 

One possibility is to use the width of $\vert \Psi \vert^2$ about the
maximum, in comparison to some $\epsilon$-dependent quantity. When the
width is sufficiently small, the certainty assumption is assumed to
apply. Consider a self-adjoint operator $A$, its eigenvectors $\psi_i$ and
eigenvalues $\lambda_i$, $A\psi_i=\lambda_i\psi_i$, $\langle \psi_i
\vert \psi_j \rangle=\delta_{ij}$ ($i,j=1,\ldots,n$). (For the qubit
case $A$ is the spin operator and $n=2$.) For a state
$\psi=\sum_{i=1}^n c_i\psi_i$, projection operators $P_i$ satisfy
$P_i\psi=c_i\psi_i$. This gives $\langle \psi \vert P_i \vert \psi
\rangle=\vert c_i\vert^2=p_i$ and $\sum_{i=1}^n p_i=1$. Let us
consider the state of $N$ copies of $\psi$,
$\Psi=\otimes_{a=1}^N\psi^{(a)}$. The frequency operators for the
eigenvalues $\lambda_i$ are
\begin{equation}
  F_i=N^{-1}\sum_{j_1,\ldots, j_N=1}^n
  \sum_{a=1}^N\delta_{ij_a}\otimes_{b=1}^N P_{j_b}.
\end{equation}
We find
\begin{align}
  &\langle \Psi \vert F_i \vert \Psi \rangle=p_i,\\
  &\langle \Psi \vert F_i^2 \vert \Psi \rangle=N^{-1}p_i+N^{-1}(N-1)p_i^2,
\end{align}
and the variances are $(\Delta F_i)^2=N^{-1}p_i(1-p_i)$.

Consider $\psi'=\sum_{i=1}^n c'_i\psi_i$ close to $\psi$, and 
require 
\begin{equation}
(p_i-p'_i)^2< \text{min}\{(\Delta F_i)^2,(\Delta
F'_i)^2\}. 
\end{equation}
This gives
\begin{equation}
  \vert c_i-c'_i\vert^2< N^{-1}(1-p_i),
\end{equation}
which leads to
\begin{equation}
  \vert\psi-\psi'\vert^2< N^{-1}(n-1).
\end{equation}
This condition is satisfied if we require $\vert\psi-\psi'\vert^2\ll
\epsilon^2$, recalling that $N\epsilon^2 < 1$. It is natural to
identify the two states $\psi$ and $\psi'$, and consider them both
approximate eigenstates of the frequency operator.

\section{Conclusions} 

We argued that attempts to derive the Born rule, either in the Many
Worlds or Copenhagen interpretation, are unsatisfactory for systems
with only a finite number of degrees of freedom. For Many
Worlds this is a serious problem, since its goal is to account for
apparent collapse phenomena---including the Born rule for
probabilities---assuming only unitary evolution of the wavefunction.
For finite number of degrees of freedom, observers on the vast
majority of branches would not deduce the Born rule.

However, we noted that discreteness of the quantum state space, even if
extremely tiny, may restore the validity of the usual arguments. Some
may regard discreteness as a radical proposal. We might argue
that it is actually less speculative than absolute continuity,
something that can never be experimentally verified.

\begin{acknowledgments}
\section{acknowledgements}

S.~H. thanks R. Hanson and J. Hormuzdiar for stimulating his interest
in this topic. The authors thank J. Hartle and S. Weinberg for
encouraging comments. A.~Z. is particularly indebted to J. Hartle for
an enlightening discussion of the decoherent history approach as
described in Ref.~\cite{Hartle2}.  S.~H. and R.~B. are supported by the
Department of Energy under DE-FG06-85ER40224, while A.~Z. is supported
in part by the National Science Foundation under grant number PHY
04-56556.

\end{acknowledgments}


\bigskip



\baselineskip=1.6pt
\begin{thebibliography}{99}

\bibitem{Everett} H. Everett, Rev.Mod.Phys {\bf 29} 3 454 (1957)


\bibitem{Hartle} J.B. Hartle, Am. J. Phys. {\bf 36} 704 (1968)

\bibitem{DeWitt} B. S. DeWitt and N. Graham in The Many Worlds
Interpretation of Quantum Mechanics, Princeton University Press (1973).

\bibitem{Hartle2} J.B. Hartle, [gr-qc/9210006].



 \bibitem{fn1}The MW formalism is familiar to anyone who studies
quantum computation. A quantum computer is an isolated system which
evolves according to a unitary quantum algorithm during a
computation. A description of the state of the quantum computer during
the computation involves a wavefunction with many branches, and no
collapse until the final read-out of the results. MW proposes that the
entire universe should be treated like a quantum computer; a brain
only perceives what \emph{appears} to be a measurement collapse as it
becomes correlated with an outcome on a particular branch of the
wavefunction, and decoheres from other branches due to vanishing
overlap. Those who do not accept MW are either asserting that: (i)
quantum mechanics is incomplete or (ii) our brains cannot be simulated
by sub-components of a quantum computer, which do split into multiple
branches. In case (i), quantum computers will, perhaps at some
threshold of complexity, fail to operate as predicted. Case (ii) also
denies the possibility of conventional artificial intelligence, since
quantum computers can simulate classical Turing machines.

\bibitem{Farhi}
  E.~Farhi, J.~Goldstone and S.~Gutmann,
  Annals Phys.\  {\bf 192}, 368 (1989).

\bibitem{CL} S. Coleman and A. Lesniewski, unpublished.

\bibitem{others} D. Deutsch, Proc. R. Soc. Lond. {\bf A455}, 3129
(1999); H.  Barnum, C.M. Caves, J. Finkelstein, C.A. Fuchs and
R. Schack, Proc. R. Soc. Lond., {\bf A456}, 1175 (2000); W. Zurek,
Phys. Rev. Lett. {\bf 90}, 120404 (2003); Rev. Mod. Phys. {\bf 75}, 715
 (2003), Phys. Rev. A {\bf 71}, 052105 (2005); R. Hanson,
Found. Phys. {\bf 33},1129 (2003), Proc. R. Soc. Lond., {\bf A462},
1619 (2006).

\bibitem{BHZ} 
R.~V.~Buniy, S.~D.~H.~Hsu and A.~Zee,
  Phys.\ Lett.\ B {\bf 630}, 68 (2005)
  [hep-th/0508039].

\bibitem{dh}
M.~Gell-Mann and J.~B.~Hartle,
  Phys.\ Rev.\ D {\bf 47}, 3345 (1993)
  [gr-qc/9210010].

\bibitem{minlength} 
X.~Calmet, M.~Graesser, S.~D.~H.~Hsu,
Phys.~Rev.~Lett. 93:211101, (2004) [hep-th/0405033],
Int.\ J.\ Mod.\ Phys.\ D {\bf 14}, 2195 (2005) [hep-th/0505144].

\bibitem{fn2} It has
not escaped our notice that discrete Hilbert space is something of an
oxymoron, given the mathematical properties of a Hilbert
space. However, the identification of quantum state space and Hilbert
space in physics has become so strong that the two are almost
interchangeable in the minds of physicists.

\bibitem{h1}
L.~Susskind, J. Math. Phys. {\bf 36}, 6377 (1995), [hep-th/9409089];
J.~M.~Maldacena, Adv. Theor. Math. Phys. {\bf 2}, 231 (1998), [hep-th/9711200].

\bibitem{h2} G.~'t~Hooft, [gr-qc/9310026]; R.~V.~Buniy and S.~D.~H.~Hsu,
  [hep-th/0510021].


\end{thebibliography}
\end{document}